\input{aipcheck}

\documentclass[
    ,final            
  ]
  {aipproc}

\layoutstyle{6x9}

\begin{document}

\title{Metallicity of the polar disk in NGC4650A: constraints for cold accretion scenario}


\author{M. Spavone}{
  address={University of Naples ''Federico II'', Via Cinthia, I80126 Naples, Italy},
  email={spavone@na.infn.it},
}
\iftrue
\author{E. Iodice}{
  address={INAF-OAC, Via Moiariello 16, I80131 Naples, Italy},
}
\author{M. Arnaboldi}{
  address={European Southern Observatory,
Karl-Schwarzschild-Stra{\ss}e 2, D-85748 Garching bei M\"{u}nchen,
Germany},
}
\author{O. Gerhard}{
  address={Max-Plank-Institut f\"{u}r
Extraterrestrische Physik, Giessenbachstra{\ss}e, D-85741 Garching
bei M\"{u}nchen, Germany},
}
\author{R. Saglia}{
  address={Max-Plank-Institut f\"{u}r
Extraterrestrische Physik, Giessenbachstra{\ss}e, D-85741 Garching
bei M\"{u}nchen, Germany},
}
\author{G. Longo}{
  address={University of Naples ''Federico II'', Via Cinthia, I80126 Naples, Italy},
}
\fi

\begin{abstract}
We used high resolution spectra in the optical and near-infrared wavelength range to study the abundance ratios and metallicities
of the HII regions associated with the polar disk in NGC4650A, in order to put constraints on the formation of the polar disk through cold gas
accretion along a filament; this might be the most realistic way by which galaxies
get their gas. We have compared the measured metallicities for the polar structure in
NGC4650A with those of different morphological types and we have found that they
are similar to those of late-type galaxies: such results is consistent with a polar disk
formed by accretion from cosmic web filaments of external cold gas.
\end{abstract}

\maketitle


\section{Introduction}



Recently, \citet{Mac06} and \citet{Bro08} have studied the formation and evolution of a
polar disk galaxy: a long-lived polar structure will form through cold gas accretion along a filament,
extended for $\sim 1$ Mpc, into the virialized dark matter halo.
The morphology and kinematics of one simulated
object are quite similar to those observed for the polar disk galaxy NGC4650A.
This galaxy is the prototype for PRGs; the polar disk is very massive,
since the total HI mass in this component is about $10^{10}M_{\odot}$, which added to the mass of stars is
comparable with the total mass in the central spheroid.


\section{Chemical abundance determination}
- \emph{Empirical methods} -
We derived the \emph{Oxygen abundance parameter}
$R_{23} = ([OII]\lambda 3727 + [OIII]\lambda\lambda 4959+5007)/H_{\beta}$ \cite{Pag79}, and the
\emph{Sulphur abundance parameter} $S_{23} = ([SII]\lambda\lambda 6717 + 6731 +[SIII]\lambda\lambda 9069 + 9532)/H_{\beta}$
 \cite{Diaz00}. We need to use both indicators to break the degeneracy with the metallicity that affect $R_{23}$;
we calibrate the oxygen abundance through the Sulphur abundance parameter by using the relation
$12+log(O/H) = 1.53logS_{23} + 8.27$, introduced by \citet{Diaz00}.

We also used another empirical method introduced by \citet{Pil01}, that propose to use the excitation
parameter $P = R_{3}/R_{23}$ (where $R_{3}=([OIII]\lambda\lambda 4929 + 5007)/H_{\beta}$), to estimate
the oxygen abundance as a function of $P$ and $R_{23}$.
The abundances derived with both methods are consistent and we also find an absence of any
metallicity gradient along the polar structure.\\

- \emph{Direct method} -
We derived the oxygen abundance of the polar disk directly by
the estimate of the $O^{++}$ and $O^{+}$ ions electron temperature. According to \citet{Izo05} and \citet{Pil06},
$12+log(OIII/H)=f(OIII[4959+5007]/H_{\beta},t_{3})$ and
$12+log(OII/H)=f(OII[3727]/H_{\beta},t_{2},Ne)$,
where $t_{3}, t_{2}$ are the electron temperatures within the OIII and OII
zones respectively, and Ne is the electron density. The total oxygen
abundance is O/H=OIII/H+OII/H and is in good agreement
with the values derived with empirical methods.

\section{Results}
In order to test the cold accretion scenario for the formation of
polar disk galaxy NGC4650A, we estimated the oxygen abundance $12+log(O/H)$
along the polar disk, by using both the empirical and direct methods.

By comparing the abundance of the polar disk with those of
galaxies of different morphological type, but same luminosity,
NGC4650A has metallicity lower than spiral galaxy disks of
the same total luminosity.

The average metallicity for the polar disk of NGC4650A is
$Z= 0.35 Z_{\odot}$ and compared with the typical values for spiral
galaxies, turns to be lower. This value is consistent with those predicted for the formation of disks by cold
accretion processes\citep{Age09}.

Moreover, the oxygen abundance along the polar disk of NGC4650A
is constant: this suggests that the metal
enrichment is not influenced by the stellar evolution of the
central spheroid and, thus, the disk was formed later. The
absence of any metallicity gradient is also found in some other
PRGs \citep{Bros09}, while a strong gradient is observed
in spiral galaxies: this turns to be consistent with infall of
metal-poor gas from outside which is still forming the disk.








\bibliographystyle{aipproc}   

\bibliography{001mspavone_biblio.bib}

\IfFileExists{\jobname.bbl}{}
 {\typeout{}
  \typeout{******************************************}
  \typeout{** Please run "bibtex \jobname" to optain}
  \typeout{** the bibliography and then re-run LaTeX}
  \typeout{** twice to fix the references!}
  \typeout{******************************************}
  \typeout{}
 }

\end{document}